\newcommand{\CLASSINPUTinnersidemargin}{18mm}
\newcommand{\CLASSINPUToutersidemargin}{12mm}
\newcommand{\CLASSINPUTtoptextmargin}{20mm}
\newcommand{\CLASSINPUTbottomtextmargin}{25mm}
\documentclass[10pt,conference,a4paper]{IEEEtran}

\usepackage{selinput}
\SelectInputMappings{
  aacute={á},
  ntilde={ñ},
  Euro={€}
}
\usepackage[T1]{fontenc}

\usepackage{times}

\usepackage{graphicx}

\usepackage{amsmath}
\usepackage{times}
\usepackage{graphicx}
\usepackage{subfigure}
\usepackage{amsfonts}
\usepackage{amssymb}
\usepackage{cite}
\usepackage{subfigure}
\usepackage{url}
\usepackage{tabularx}
\usepackage{array}
\newcolumntype{Y}{>{\hsize=0.7\hsize\centering\arraybackslash}X}
\newcolumntype{Z}{>{\hsize=1.3\hsize\centering\arraybackslash}X}

\DeclareGraphicsExtensions{.png,.eps,.ps,.pdf}

\begin{document}

\title{Calibration electronics for the 30 and 40 GHz instrument (TFGI) of the QUIJOTE experiment}


\author{
\authorblockN{Jorge Luis Díaz-Acosta$^{(1)}$, Roger John Hoyland$^{(2)}$, Francisco Javier Casas$^{(1)}$, Enrique Martinez-González$^{(1)}$,\\on behalf of the QUIJOTE Team}
\authorblockA{dacosta@ifca.es, rjh@iac.es, casas@ifca.es, martinez@ifca.es}
\authorblockA{$^{(1)}$Observational Cosmology and Instrumentation. Instituto de Física de Cantabria. Edificio Juan Jordá, Campus de la \\
University of Cantabria, Avenida de los Castros s/n, 39005. Santander (Cantabria).}
\authorblockA{$^{(2)}$Electronics Department. Instituto de Astrofísica de Canarias. Calle de la Vía Láctea s.n. \\
38205 La Laguna (Tenerife).}
}

\maketitle

\begin{abstract}
The 30 and 40 GHz instruments of the QUIJOTE radio astronomy experiment use very sensitive receivers which need to be characterized both to ensure their proper functioning and for their calibration. Given the age of the PXI-1031DC device used up until now for this purpose, the design of a new device, the Calibrator TFGI QUIJOTE, has been proposed. This new device has been designed with the idea of being simple and modular, easy to use and upgrade. It is built on a Raspberry Pi 5 system, using an MCC 118 as analog reading module. This system, widely known among both makers and professionals, not only meets the requirements of modularity and ease of use, but also ensures that the device is much more economically competitive than the PXI. A first version of the device has been built and tested, obtaining good results in accuracy and ease of use.
\end{abstract}

\section{Introduction}
\label{cap:introduction}

The QUIJOTE experiment \cite{ref:Quijote_CMB} is an international collaboration between the Instituto de Astrofísica de Canarias (IAC), the Instituto de Física de Cantabria (IFCA), the Universities of Cantabria (UC), Manchester and Cambridge, and the company IDOM, which involves the design, manufacture and commissioning of several instruments to be installed in radio telescopes at the Teide Observatory on the island of Tenerife. Its objective is the characterization of the Cosmic Microwave Background (CMB) and other galactic and extragalactic emission processes in the range of 10 to 47 GHz.

\begin{figure*}[t]
\centering
\subfigure{
    \includegraphics[width=0.75\textwidth]{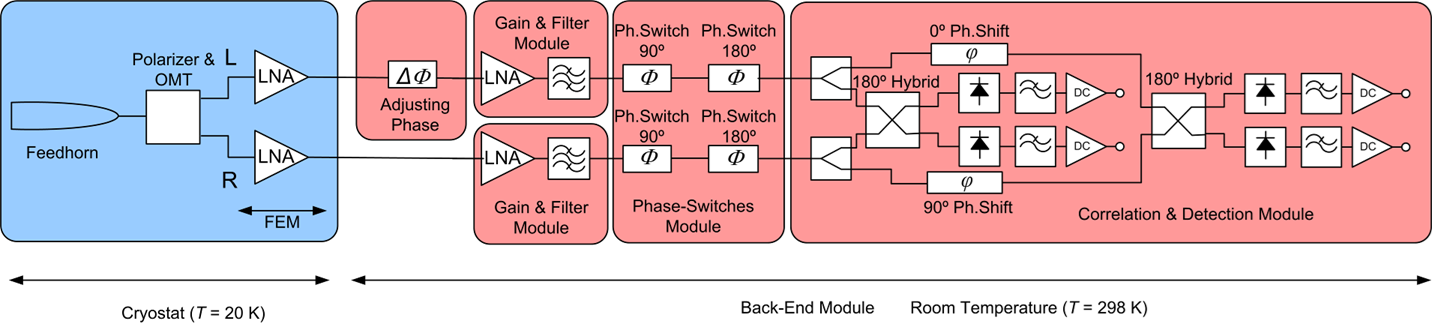}
}
\subfigure{
    \includegraphics[width=0.75\textwidth]{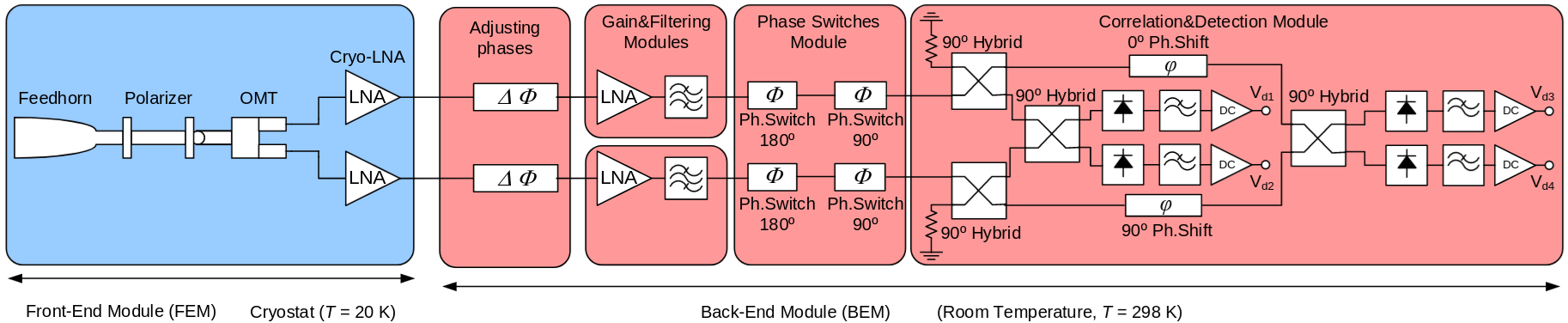}
}
\caption{Diagram of the TFGI instrument receivers (pixels). Top: TGI Instrument. Bottom: FGI Instrument. Figure published in \cite{ref:fasano2026quijotetfgipolarizationcalibration}}
\label{fig:esquema_TFGI}
\end{figure*}

Two of these instruments are the Thirty Gigahertz Instrument (TGI) and the Forty Gigahertz Instrument (FGI) \cite{ref:TFGI_URSI_2016}, designed for the detection of the so-called ``B-modes'' of the CMB around the frequencies of 30 and 40 GHz, respectively. These instruments are made up of a low-noise Front-End Module (FEM), responsible of receiving and conditioning the signal into two signals corresponding to left (L) and right (R) polarization, followed by a Back-End Module (BEM), in charge of the correlation of the signals and their direct detection at low frequency. In addition, the BEM is also designed to perform a phase modulation of the signals for the correction of systematic errors of the receiver by modulating the polarization of the input signal. A diagram of the TGI and FGI receivers can be seen in Fig. \ref{fig:esquema_TFGI}. The TGI has 30 of these receivers in one BEM, while the FGI has 10 receivers.

Given that the CMB polarization is low, the receivers of the BEMs of these instruments are required to be very sensitive to guarantee the quality of the obtained data. Therefore, it is important to have a tool that allows characterizing their output for known input signals, both to check their proper functioning and to be able to calibrate their response during the instrument's commissioning.

\begin{figure}[t]
\centering
\includegraphics[width=0.75\columnwidth]{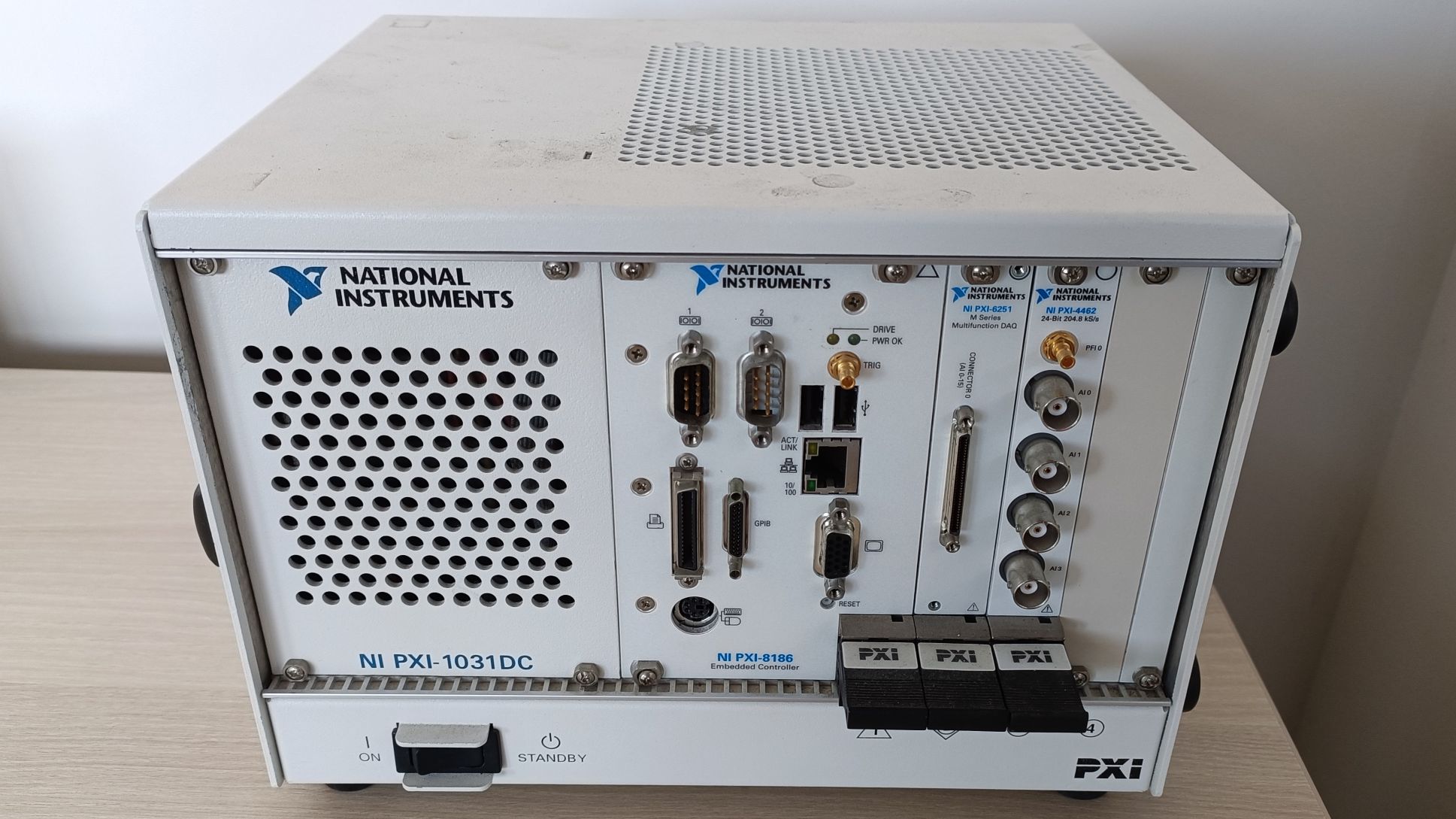}
\caption{{PXI}-1031DC platform supplied by DICOM-UC}
\label{fgi:pxi-1031dc}
\end{figure}

Until now, a PXI-1031DC had been used for this task, a PC-based modular test platform developed by National Instruments \cite{ref:PXI_website} and oriented towards the automation of data acquisition processes. This has a PXI-8186 embedded controller installed as its core, in addition to incorporating a PXI-4462 analog input module for reading the output of the receivers and a PXI-6251 multifunction I/O module for switching the states of the phase modulator of the receivers. The equipment runs on a Windows XP operating system, and the control of the switching of these states and the data visualization was carried out through a program developed in LabVIEW 2012. An image of the equipment can be seen in Fig. \ref{fgi:pxi-1031dc}. However, given the steep learning curve of this program, and the impossibility of updating it to a more modern version of the operating system to have access to more modern alternatives (both due to hardware limitations and the closed environment of the equipment), it has been determined necessary to replace it with a new, more modern system.

After discarding the option of acquiring a more modern PXI due to its high economic cost, the alternative of an in-house system has been proposed, called QUIJOTE TFGI Calibrator, developed entirely at the IAC using modern, accessible hardware adapted to the needs of the instrument. This equipment has been tested during the last repair and commissioning campaign of the instrument, obtaining good results in terms of ease of use and precision.

\section{QUIJOTE TFGI Calibrator}
\label{cap:calibrador}

The QUIJOTE TFGI Calibrator is a device conceived as an easy-to-use and upgradeable measurement and analysis tool. Its initial design is optimized to characterize the pixel behavior of the TGI and FGI instruments, although its modular design would allow adapting it to other instruments with similar operation with minimal changes.

\begin{figure}[t]
\centering
\subfigure{
    \includegraphics[width=0.75\columnwidth]{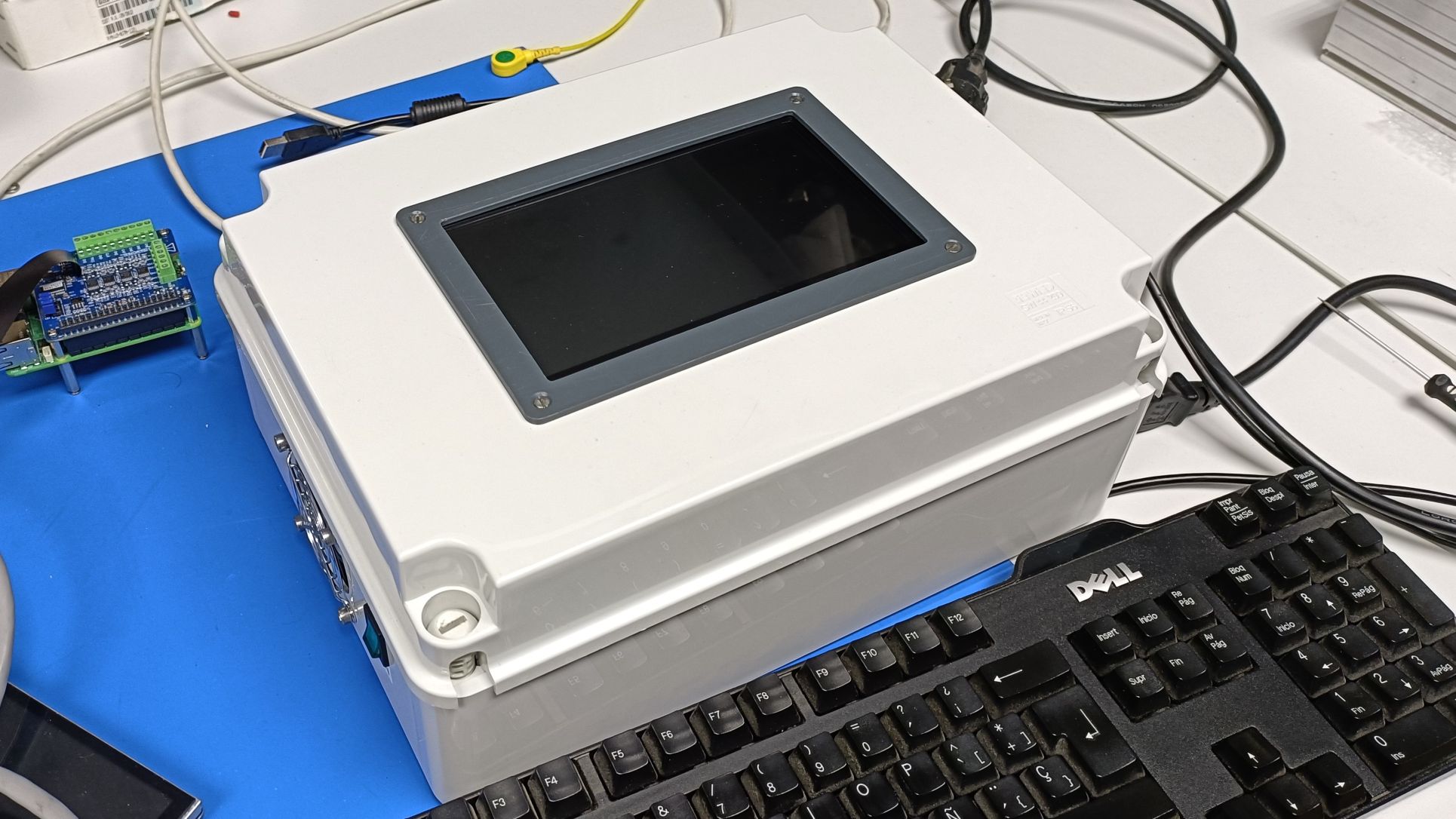}
}
\subfigure{
    \includegraphics[width=0.75\columnwidth]{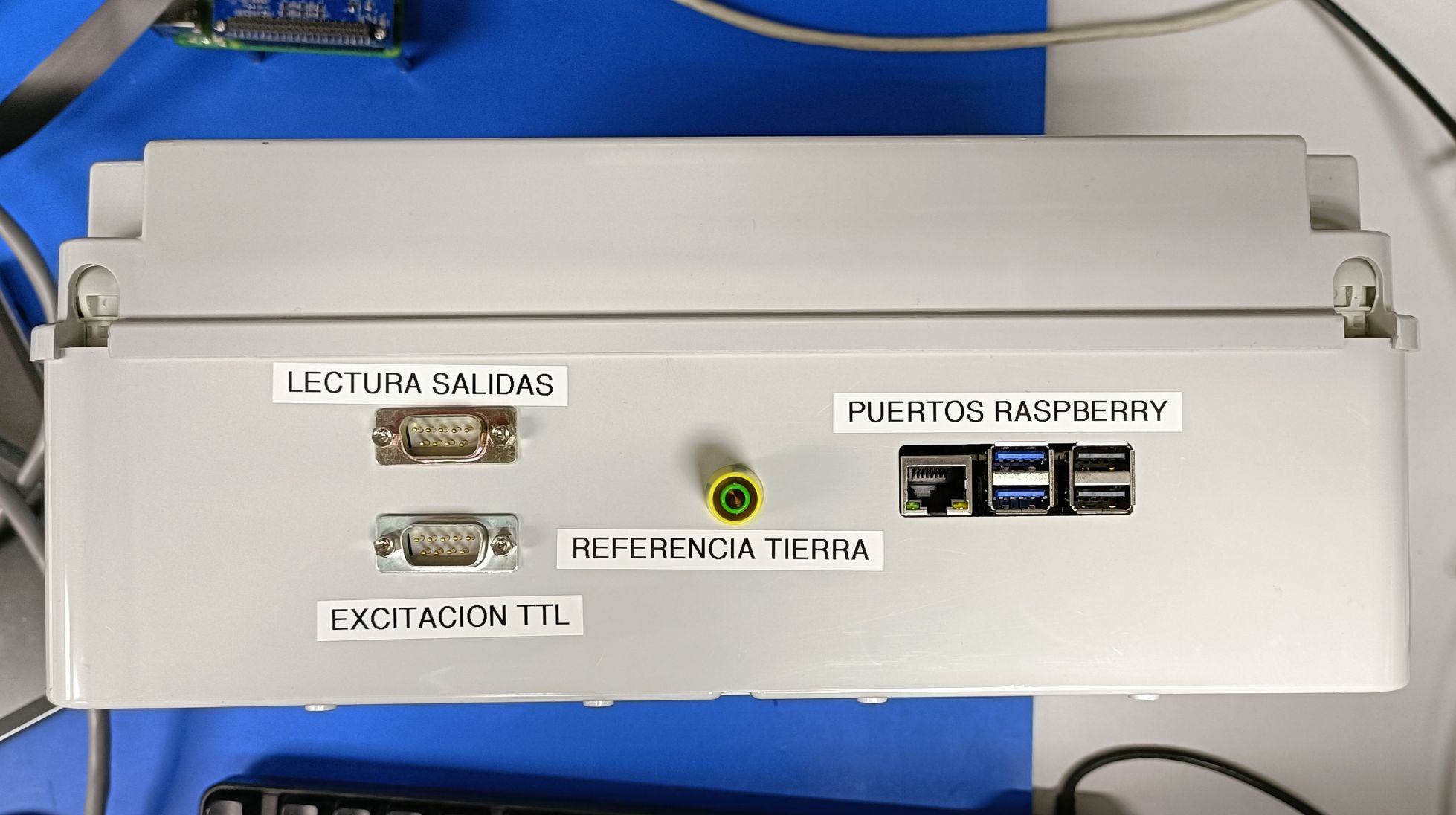}
}
\caption{Exterior of the QUIJOTE TFGI Calibrator. Top: Top panel. Bottom: Rear panel}
\label{fig:calibrator_exterior}
\end{figure}

The device consists of a simple box with a touch screen on the top and connectors on the rear panel, some dedicated to reading the outputs of the instruments, connecting the control signals of the phase modulators of the measured pixel and ground reference, and other general ones for peripherals. In addition, the device is designed to be able to connect and communicate with any network analyzer device compatible with the VXI-11 communications protocol, allowing it to accurately control both the power and frequency of the signal injected into the characterized pixel during the test. The exterior appearance of the device can be seen in Fig. \ref{fig:calibrator_exterior}.

\subsection{Device hardware}
\label{subcap:calibrator_hardware}

At the hardware level, the device has been designed with accessibility and ease of expansion and improvement in mind. The core of the equipment is a Raspberry Pi 5, the most recent version of the platform, widely used worldwide for both ``maker'' and professional projects, and which has extensive documentation freely available on the internet and an equally large community. Table \ref{tab:Raspberry_specs} shows its main specifications.

\begin{table}[t]
\renewcommand{\arraystretch}{1.3}
\caption{Raspberry Pi 5 specifications \cite{ref:raspberry_pi}}
\label{tab:Raspberry_specs}
\begin{center}
\begin{tabularx}{\columnwidth}{|Y|Z|}
\hline
\textbf{CPU} & 64-bit quad-core Arm Cortex-A76 at 2.4GHz\\
\hline
\textbf{GPU} & VideoCore VII\\
\hline
\textbf{RAM} & 8GB LPDDR4X-4267 SDRAM\\
\hline
\textbf{Ports} & 2xUSB3.0, 2xUSB2.0, Ethernet with PoE+ support, 2x4-lane MIPI\\
\hline
\textbf{Expansion ports} & 40-pin GPIO header\\
\hline
\textbf{Memory} & microSD with support for SDR104 fast read mode\\
\hline
\end{tabularx}
\end{center}
\end{table}

For data acquisition, an MCC 118 DAQ HAT (Hardware Attached on Top) board, developed by Digilent specifically for Raspberry Pi devices, has been installed. This board is designed to measure single-ended voltages in a range of $\pm$10V with a 12-bit resolution on up to 8 simultaneous channels. It can also perform sampling sweeps at rates of up to 100kS/s, with a trigger pin \cite{ref:MCC188_specs} that allows the synchronization of the sweeps with the switching sequence of the instrument's phase modulator. Although the board is not capable of working with differential voltages by default, this deficiency was solved by software later on.

\begin{figure}[t]
\centering
\includegraphics[width=0.75\columnwidth]{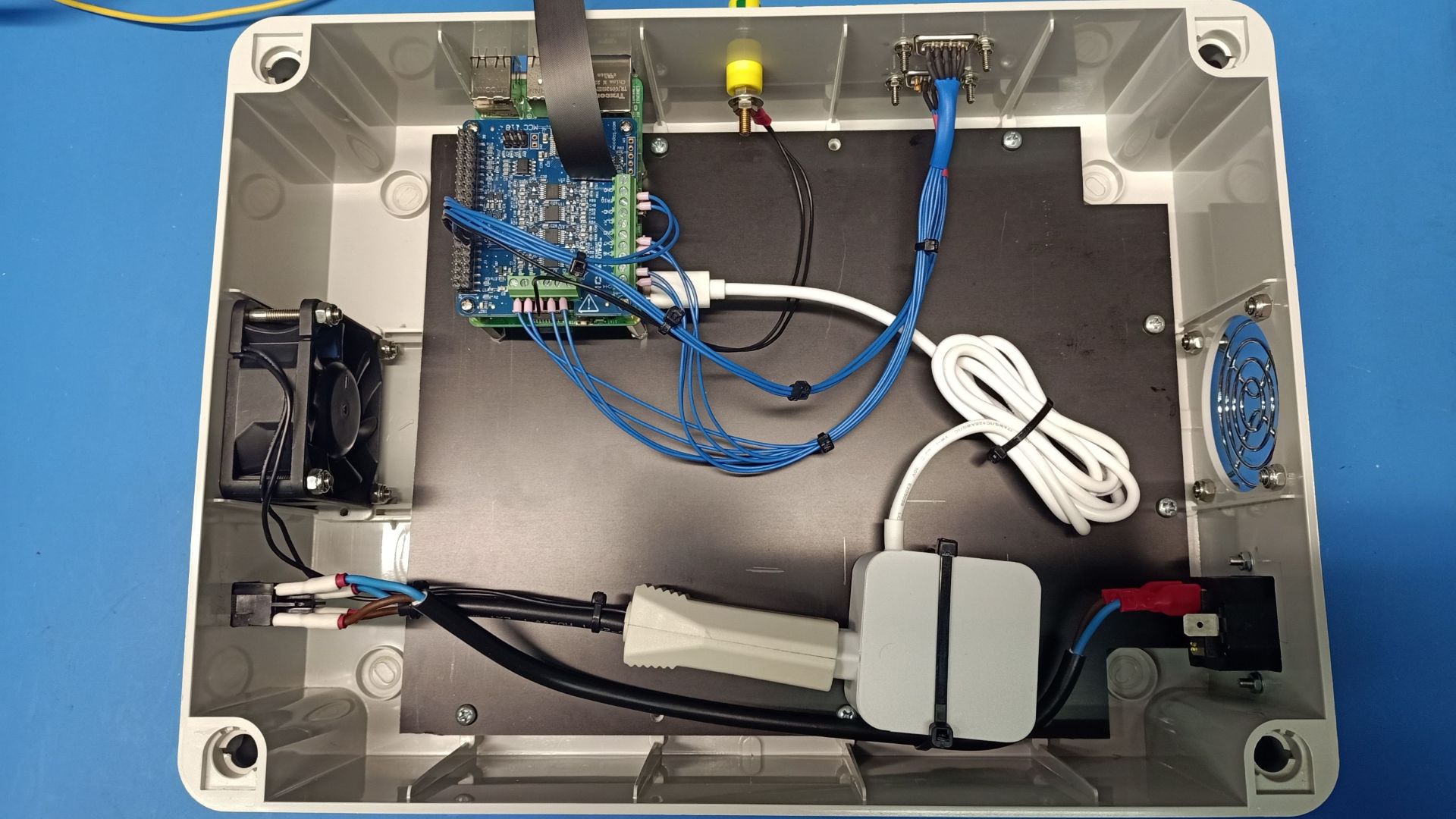}
\caption{Interior of the QUIJOTE TFGI Calibrator}
\label{fig:calibrator_interior}
\end{figure}

For the control of the phase modulator, free pins of the Raspberry Pi header have been allocated. Both the channels of the MCC 118 and the control pins have been routed to 9-pin D-sub connectors on the rear panel of the device. A reference port has also been added to be able to match the ground level of the instrument to the ground level of the device and guarantee that the measurements are correct. Finally, a DFRobot DFR0678 LCD touch screen with MIPI connector has been used as a human-machine interface for the device. The interior of the finished device can be seen in Fig. \ref{fig:calibrator_interior}.

\subsection{Device software}

The operating software of the device follows the same philosophy of ease of use and expansion. The device runs on the Raspbian GNU/Linux operating system, Raspberry Pi's own Linux distribution, open-source and widely used and documented on the internet.

\begin{figure}[t]
\centering
\includegraphics[width=0.75\columnwidth]{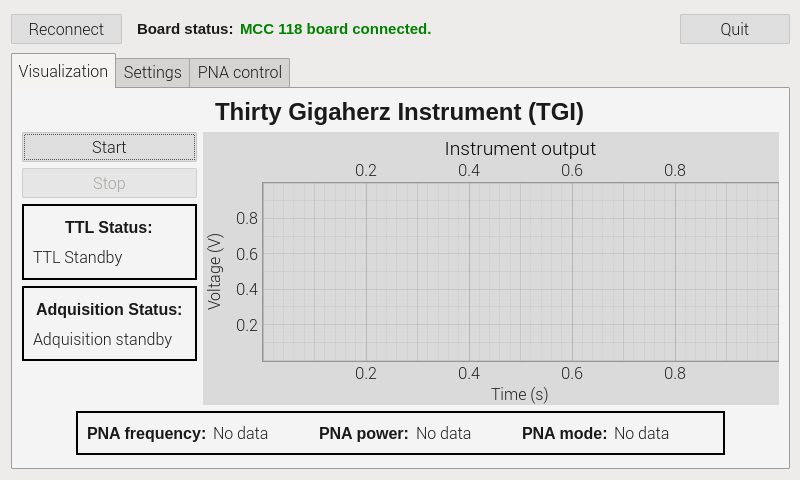}
\caption{Main page of the QUIJOTE TFGI Calibrator interface}
\label{fig:calibrator_interface}
\end{figure}

In turn, the program that serves as the user interface of the device is fully implemented in Python, a widely known and used language, which facilitates its future updating and improvement. Communication with the MCC 118 is done using the open-source Daqhat library, developed and maintained by the board manufacturer itself, while the interface has been implemented using PyQt5, a Python wrapper for the Qt library. A screenshot of the main page of the interface can be seen in Fig. \ref{fig:calibrator_interface}.

\section{Comparison between PXI and QUIJOTE TFGI Calibrator}

Below is a comparison of the QUIJOTE TFGI Calibrator with the PXI-1031DC.

\subsection{Capabilities}

In section \ref{subcap:calibrator_hardware}, the technical hardware characteristics of the QUIJOTE TFGI Calibrator were described. If we compare the capabilities of the Raspberry Pi with those of the PXI-8186 that we can see in Table \ref{tab:PXI-8186_specs}, we can verify that the Raspberry Pi has superior characteristics to the PXI, since it not only has more power, but the use of the Raspberry's GPIO allows not needing an external I/O module, as the PXI does require.

\begin{table}[t]
\renewcommand{\arraystretch}{1.3}
\caption{PXI-8186 specifications \cite{ref:PXI8186specs}}
\label{tab:PXI-8186_specs}
\begin{center}
\begin{tabularx}{\columnwidth}{|Y|Z|}
\hline
\textbf{CPU} & Pentium 4-M at 2.2GHz\\
\hline
\textbf{GPU} & Intel Extreme Graphics\\
\hline
\textbf{RAM} &  256 MB DDR\\
\hline
\textbf{Hard drive} & 30GB\\
\hline
\textbf{Ports} & 2xserial ports, 2xUSB2.0, Ethernet 100 BaseTX, GPIB interface (IEEE 488.2)\\
\hline
\end{tabularx}
\end{center}
\end{table}

However, the PXI-4462 analog input module does represent an important advantage of the PXI with respect to the QUIJOTE TFGI Calibrator. This module can also measure voltages in the $\pm$10V range, and can perform sampling sweeps at speeds up to 204.8kS/s with a 24-bit resolution \cite{ref:PXI4462specs}, more than double what the MCC 118 board can offer. Furthermore, its four ports are configured to measure differential voltages, which is relevant for the TFGI instrument, since its outputs are differential.

\begin{figure}[t]
\centering
\includegraphics[width=0.85\columnwidth]{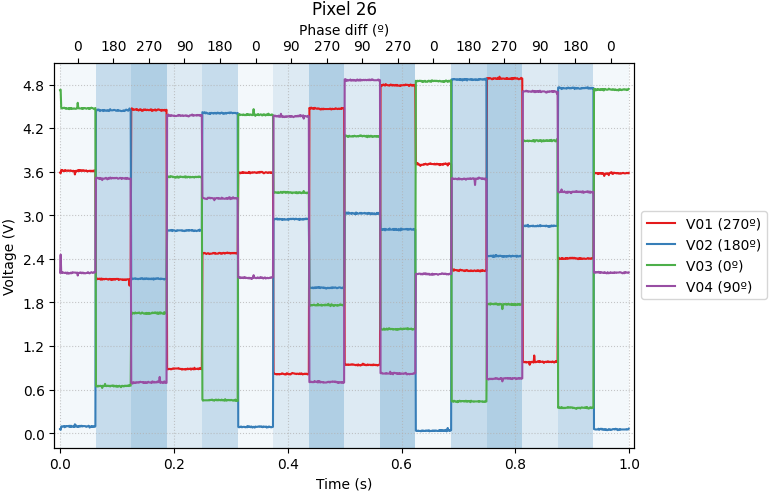}
\caption{Reading of the TGI pixel 26 output to a polarized signal of 31 GHz and -50 dBm generated with the PNAx N5245B obtained with data captured by the QUIJOTE TFGI Calibrator}
\label{fig:ejemplo_pixel}
\end{figure}

Despite the above, the MCC 118 board maintains the advantage if the instrument uses single-ended voltage outputs, where the board would have up to 8 ports, compared to the 4 that the PXI-4462 would maintain, since these only work in differential mode. Moreover, given the output voltage levels with which the instrument normally works, the difference in accuracy between both acquisition boards becomes irrelevant. In Fig. \ref{fig:ejemplo_pixel} an example of a characterized signal can be seen using the QUIJOTE TFGI Calibrator for the four outputs (Vd1, Vd2, Vd3, and Vd4) of the TGI instrument when it receives a pure Q-polarized signal at -50 dBm at the center frequency of the instrument's central band (31 GHz).

\begin{table*}[t]
\renewcommand{\arraystretch}{1.3}
\caption{Cost comparison of a new PXI and the QUIJOTE TFGI Calibrator\textsuperscript{1}}
\label{tab:Comparacion_PXI_Calibrador}
\begin{center}
\begin{tabular}{|c|c|c|c|}
\hline
\multicolumn{2}{|c|}{\textbf{PXI}} & \multicolumn{2}{c|}{\textbf{Calibrator}} \\
\hline
\textbf{Component} & \textbf{Price} & \textbf{Component} & \textbf{Price} \\
\hline
PXI-8822 Core & 3698 € & Raspberry Pi 5 Core & 106.48 € \\
\hline
PXIe-4464 Analog input module &  7993 € & MCC 118 modules & 84.77 € \\
\hline
PXI-6509 I/O module &  937 € & Raspberry power cable & 14.97 € \\
\hline
NI PXIe-1071 Chassis & 1957 € & Case and other components & 95€ \\
\hline
\textbf{Subtotal} & \textbf{14585 €} & \textbf{Subtotal} & \textbf{304.18 €}\\
\hline
\end{tabular}
\end{center}
\begin{center}
\parbox{0.65\textwidth}{\footnotesize{\textsuperscript{1}Average reference prices in official suppliers as of March 2026}}
\end{center}
\end{table*}

\subsection{Ease of use}

As already mentioned in section \ref{cap:introduction}, the PXI-8186 controller of the PXI-1031DC runs on Windows XP. This is a version with proprietary National Instruments drivers and software and cannot be updated both for this reason and because of the age of the device, which limits the options for developing applications for the characterization of the instrument to the use of the LabVIEW 2012 program, which is not only an old version of the program, but also a program with a steep learning curve and not widely known outside certain fields.

In contrast, the QUIJOTE TFGI Calibrator, being implemented on a Linux-based device, all its software is open-source and easily upgradeable, besides being compatible with modern technologies and languages such as Python, which facilitates modifications and future improvements to the device. Furthermore, the interface currently implemented in the device was designed with simplicity and ease of use in mind, allowing even users not trained in the equipment's design to use it without complications.

Weight is another point in favor of the QUIJOTE TFGI Calibrator in terms of facilitating its handling. With all its modules installed, the total weight of the PXI-1031DC exceeds 6 kg \cite{ref:PXI8186specs}\cite{ref:PXI4462specs}\cite{ref:PXI1031specs}, which makes it difficult to handle comfortably when having to move it to the instrument's location. On the other hand, once assembled, the Calibrator is estimated to weigh less than 1 kg, which allows comfortable one-handed handling. Additionally, its touch screen means that, if necessary, no extra peripherals are required for viewing and control, whereas the PXI requires the use of a screen, keyboard, and mouse for local use. 

\subsection{Cost}

One of the biggest advantages of the QUIJOTE TFGI Calibrator over the PXI-1031DC lies in the cost of these devices. Many of the mentioned flaws and shortcomings of the PXI with respect to the Calibrator can be solved by acquiring a more modern version of it. However, the high cost of these devices makes it unattractive compared to the Calibrator. As can be seen in Table \ref{tab:Comparacion_PXI_Calibrador}, the acquisition cost of just a single PXI component is already higher than the total cost of the QUIJOTE TFGI Calibrator, which makes the device much more economically competitive than the PXI.

\section{Conclusions}

The QUIJOTE TFGI Calibrator proves to be a solid alternative to traditional PXI systems for use in the characterization and calibration of radio astronomy instruments such as the TFGI. Its simple and modular design, based on open-source hardware and software widely used in both ``maker'' and professional environments, guarantees not only its ease of use and the quality of the obtained results, but also allows its easy adaptation to other systems and the implementation of improvements.

Currently, after this initial version, several improvements are already planned, such as a controller for rotators that allows controlling the rotation of the horns used as emitters during the instrument's characterization and calibration procedures, which would allow controlling the polarization of the input signal to the instrument, in addition to the implementation of an option for performing automatic sweeps in both frequency and power.


\section*{Acknowledgements}

The authors acknowledge the funding of this work, obtained through the research projects with reference PID2022-139223OB-C21 and with European funds "European Union NextGenerationEU/PRTR". Partial funding from the Ministry of Science, Innovation and Universities, projects: AYA2017-84185-P, IACA15-BE-3707, EQC2018-004918-P and the FEDER Agreement INSIDE-OOCC (ICTS-2019-03-IAC-12) is also acknowledged. We also acknowledge the support of the Severo Ochoa programs SEV-2015-0548 and CEX2019-000920-S. Thanks also to the support obtained from the Department of Communications Engineering of the University of Cantabria (DICOM-UC). Finally, we thank Airam Marcos-Caballero for his help in formatting with \LaTeX.


\bibliography{bibliography_english_version}

\end{document}


\title{Electrónica de calibración para el instrumento de 30 y 40 GHz (TFGI) del experimento QUIJOTE}


\author{
\authorblockN{Jorge Luis Díaz-Acosta$^{(1)}$, Roger John Hoyland$^{(2)}$, Francisco Javier Casas$^{(1)}$, Enrique Martinez-González$^{(1)}$,\\on behalf of the QUIJOTE Team}
\authorblockA{dacosta@ifca.es, rjh@iac.es, casas@ifca.es, martinez@ifca.es}
\authorblockA{$^{(1)}$Cosmología Observacional e Instrumentación. Instituto de Física de Cantabria. Edificio Juan Jordá, Campus de la \\
Universidad de Cantabria, Avenida de los Castros s/n, 39005. Santander (Cantabria).}
\authorblockA{$^{(2)}$Departamento de Electrónica. Instituto de Astrofísica de Canarias. Calle de la Vía Láctea s.n. \\
38205 La Laguna (Tenerife).}
}

\maketitle

\begin{abstract}
The 30 and 40 GHz instruments of the QUIJOTE radio astronomy experiment use very sensitive receivers which need to be characterized both to ensure their proper functioning and for their calibration. Given the age of the PXI-1031DC device used up until now for this purpose, the design of a new device, the Calibrator TFGI QUIJOTE, has been proposed. This new device has been designed with the idea of being simple and modular, easy to use and upgrade. It is built on a Raspberry Pi 5 system, using a MCC 118 as analog reading module. This system, widely known among both makers and professionals, not only meets the requirements of modularity and ease of use, but also ensures that the device is much more economically competitive than the PXI. A first version of the device has been built and tested, obtaining good results in accuracy and ease of use.
\end{abstract}

\section{Introducción}
\label{cap:introduction}

El experimento QUIJOTE \cite{ref:Quijote_CMB} es una colaboración internacional entre el Instituto de Astrofísica de Canarias (IAC), el Instituto de Física de Cantabria (IFCA), las Universidades de Cantabria (UC), Manchester y Cambridge, y la empresa IDOM, que conlleva el diseño, la fabricación y puesta en funcionamiento de varios instrumentos a instalar en radiotelescopios del Observatorio de El Teide en la isla de Tenerife. Su objetivo es la caracterización del Fondo Cósmico de Microondas (CMB, de sus siglas en inglés) y otros procesos de emisión galácticos y extragalácticos en el rango de 10 a 47 GHz.

\begin{figure*}[t]
\centering
\subfigure{
    \includegraphics[width=0.75\textwidth]{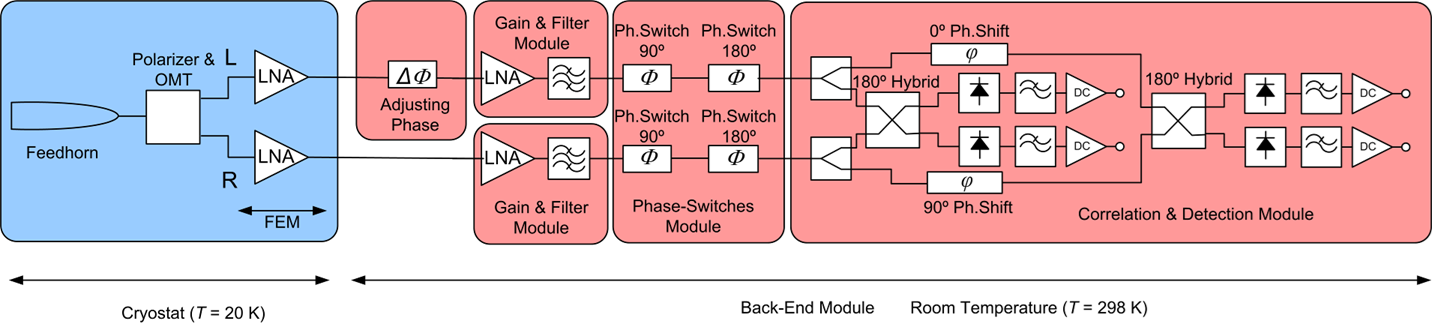}
}
\subfigure{
    \includegraphics[width=0.75\textwidth]{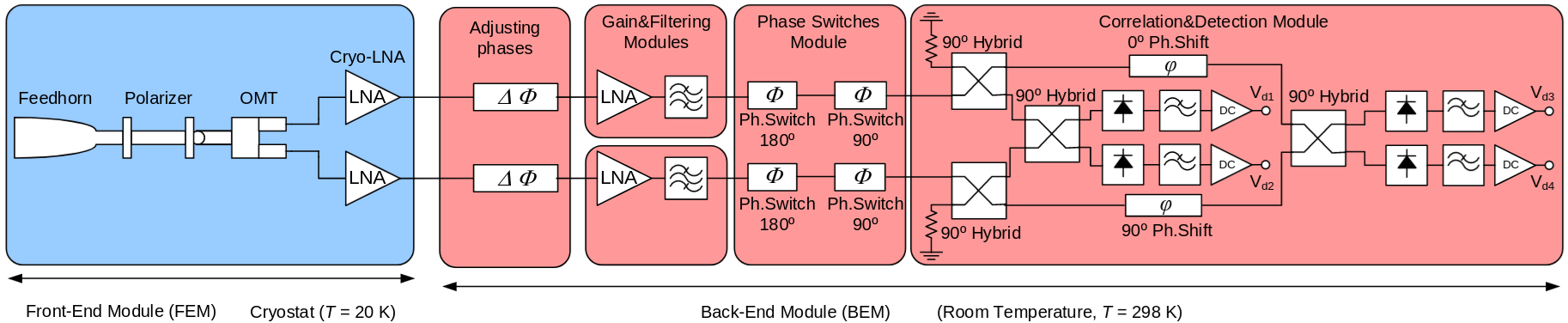}
}
\caption{Esquema de los receptores (píxeles) del instrumento TFGI. Arriba: Instrumento TGI. Abajo: Instrumento FGI. Figura publicada en \cite{ref:fasano2026quijotetfgipolarizationcalibration}}
\label{fig:esquema_TFGI}
\end{figure*}

Dos de estos instrumentos son el Thirty Gigahertz Instrument (TGI) y el Forty Gigahertz Instrument (FGI) \cite{ref:TFGI_URSI_2016}, diseñados para la detección de los denominados ``modos-B'' del CMB en torno a las frecuencias de 30 y 40 GHz, respectivamente. Estos instrumentos están conformados por un módulo frontal de bajo ruido, Front-End Module (FEM), encargado de recepción y acondicionamiento de la señal en dos señales correspondientes a la polarización a izquierdas (L) y a derechas (R), seguido de un módulo posterior, Back-End Module (BEM), encargado de la correlación de las señales y su detección directa a baja frecuencia. Además, el BEM también se encarga de realizar una modulación de fase de las señales para la corrección de errores sistemáticos del receptor mediante la modulación de la polarización de la señal de entrada. En la fig. \ref{fig:esquema_TFGI} puede verse un esquema de los receptores del TGI y del FGI. El TGI cuenta con 30 de estos receptores en un BEM, mientras que el FGI cuenta con 10 receptores.

Dado que la polarización del CMB es baja, se requiere que los receptores de los BEMs de estos instrumentos sean muy sensibles para poder garantizar la calidad de los datos obtenidos. Por ello, es importante contar con una herramienta que permita caracterizar su salida para señales de entrada conocidas, tanto para comprobar su buen funcionamiento como para poder calibrar su respuesta durante las puestas a punto del instrumento.

\begin{figure}[t]
\centering
\includegraphics[width=0.75\columnwidth]{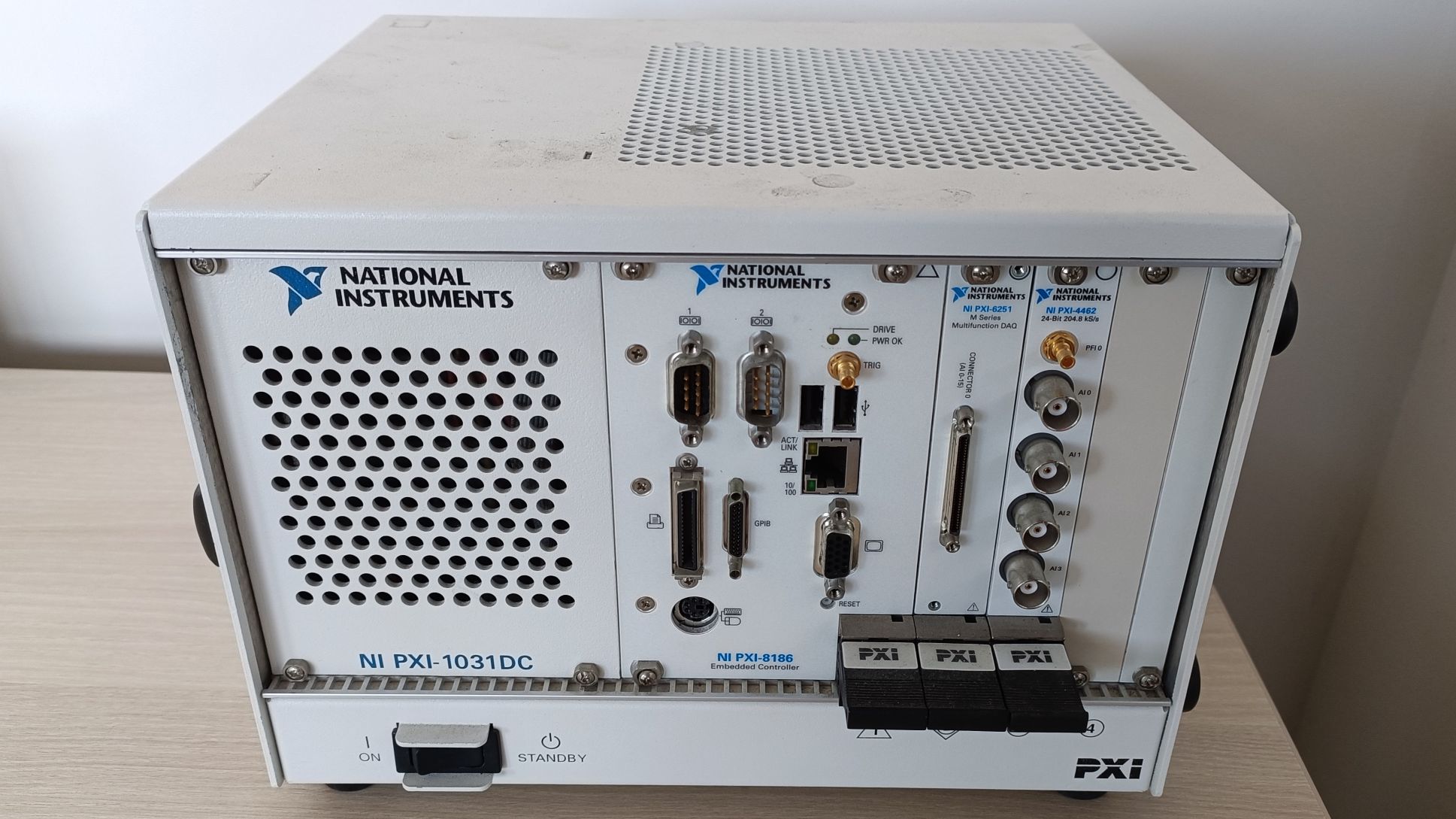}
\caption{Equipo {PXI}-1031DC suministrado por el DICOM-UC}
\label{fgi:pxi-1031dc}
\end{figure}

Hasta ahora, para esta tarea se había hecho uso de un PXI-1031DC, una plataforma de pruebas modular basada en PC desarrollada por National Instruments \cite{ref:PXI_website} y orientada a la automatización de procesos de adquisición de datos. Este tiene instalado como núcleo el controlador embebido PXI-8186, además de incorporar un módulo de lectura analógica PXI-4462 para la lectura de la salida de los receptores y un módulo de E/S multifunción PXI-6251 para la conmutación de los estados del modulador de fase de los receptores. El equipo funciona sobre un sistema operativo Windows XP, y el control de la conmutación de estos estados y la visualización de los datos se realizaba a través de un programa desarrollado en LabVIEW 2012. Puede verse una imagen del equipo en la fig. \ref{fgi:pxi-1031dc}. Sin embargo, dada la pronunciada curva de aprendizaje de este programa, y la imposibilidad de actualizarlo a una versión más moderna del sistema operativo para tener acceso a alternativas más modernas (tanto por limitaciones de hardware como del entorno cerrado del equipo), se ha determinado necesario su sustitución por un nuevo equipo más moderno.

Tras descartarse la opción de adquirir un PXI más moderno debido a su alto coste económico, se ha propuesto la alternativa de un sistema propio, denominado Calibrador {TFGI QUIJOTE}, desarrollado íntegramente en el IAC usando hardware moderno, accesible y adaptado a las necesidades del instrumento. Este equipo ha sido probado durante la última campaña de reparación y puesta a punto del instrumento, obteniéndose buenos resultados en facilidad de uso y precisión.

\section{Calibrador TFGI QUIJOTE}
\label{cap:calibrador}

El calibrador TFGI QUIJOTE es un dispositivo concebido como una herramienta de medida y análisis fácil de usar y mejorar. Su diseño inicial está optimizado para caracterizar el comportamiento de los píxeles de los instrumentos TGI y FGI, aunque su diseño modular permitiría adaptarlo a otros instrumentos de funcionamiento similar con cambios mínimos.

\begin{figure}[t]
\centering
\subfigure{
    \includegraphics[width=0.75\columnwidth]{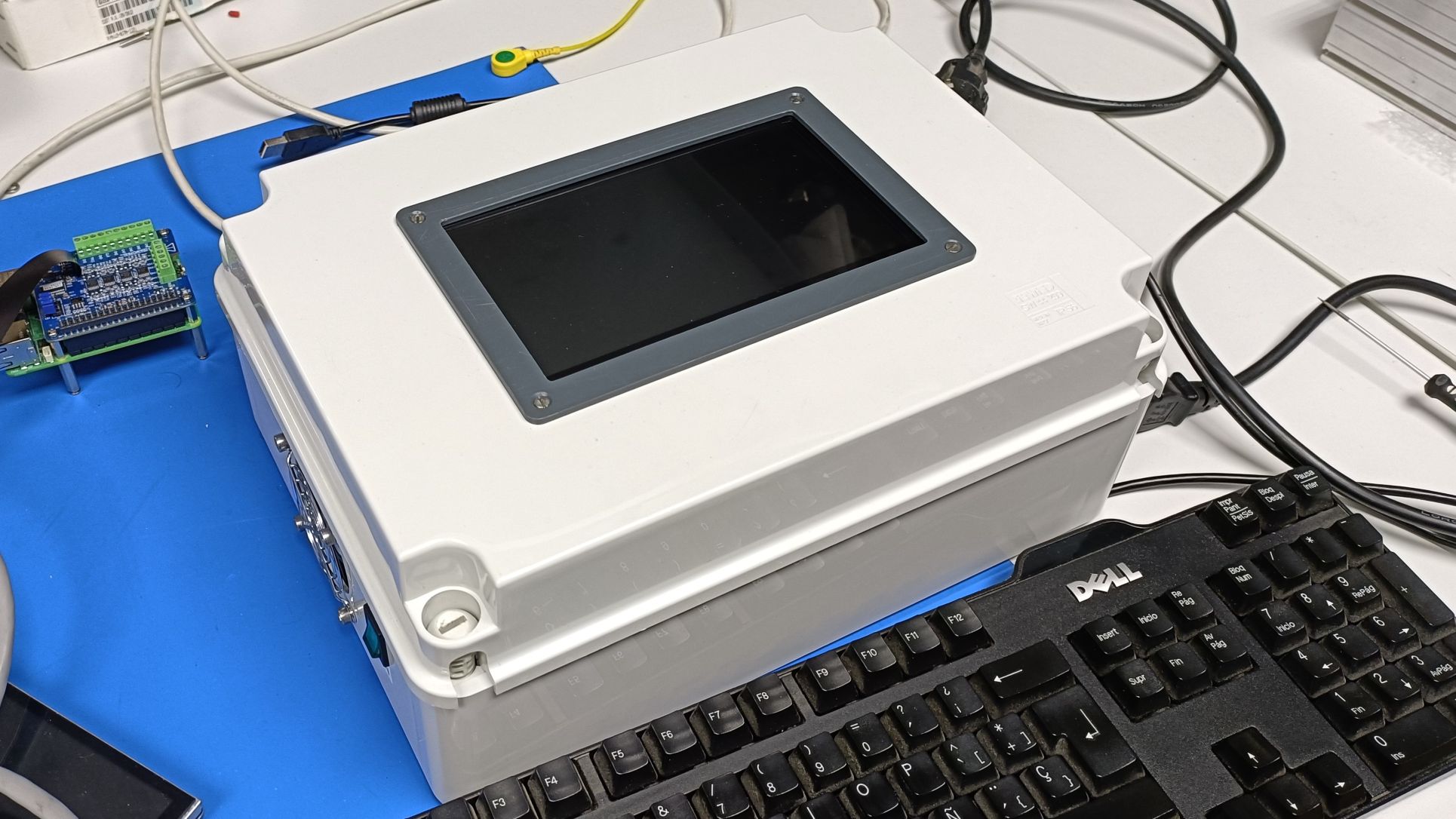}
}
\subfigure{
    \includegraphics[width=0.75\columnwidth]{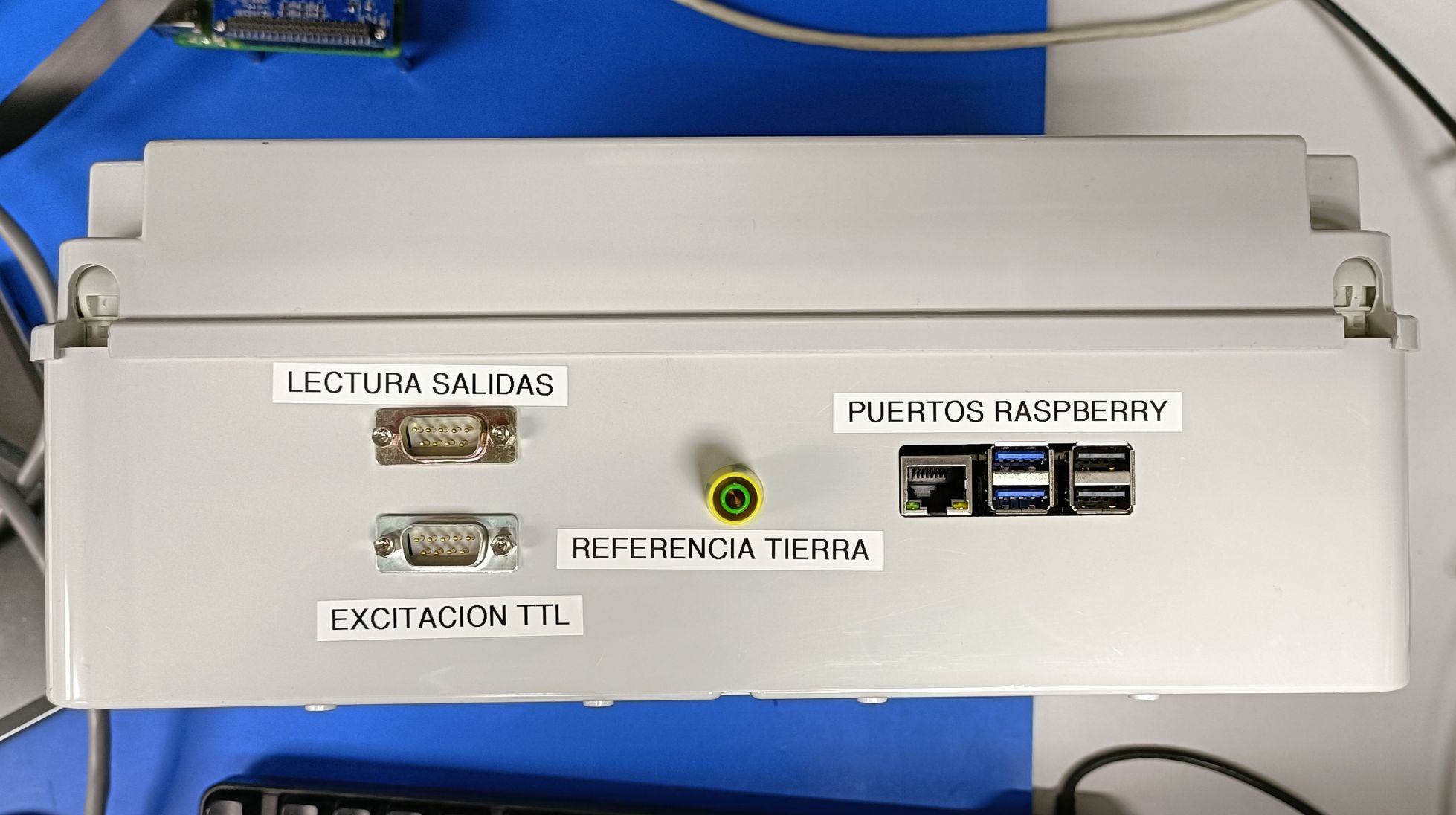}
}
\caption{Exterior del calibrador TFGI QUIJOTE. Arriba: Panel superior. Abajo: Panel trasero}
\label{fig:calibrator_exterior}
\end{figure}

El dispositivo está conformado por una caja sencilla con una pantalla táctil en la parte superior y conectores en el panel trasero, algunos dedicados para la lectura de las salidas de los instrumentos, la conexión de las señales de control de los moduladores de fase del píxel medido y referencia de tierra, y otros generales para periféricos. Además, el dispositivo está diseñado para poder conectarse y comunicarse con cualquier dispositivo analizador de redes compatible con el protocolo de comunicaciones VXI-11, permitiéndole controlar con precisión tanto la potencia como la frecuencia de la señal inyectada al píxel caracterizado durante la prueba. Puede observarse el aspecto exterior del dispositivo en la  fig. \ref{fig:calibrator_exterior}.

\subsection{Hardware del dispositivo}
\label{subcap:calibrator_hardware}

A nivel de hardware, el dispositivo ha sido diseñado con la accesibilidad y facilidad de ampliación y mejora en mente. El núcleo del equipo es una Raspberry Pi 5, versión más reciente de la plataforma, utilizada ampliamente en todo el mundo para proyectos tanto ``maker'' como profesionales, y que cuenta con una amplia documentación disponible libremente en internet y una igualmente amplia comunidad. La tabla \ref{tab:Raspberry_specs} muestra sus principales especificaciones.

\begin{table}[t]
\renewcommand{\arraystretch}{1.3}
\caption{Especificaciones Raspberry Pi 5 \cite{ref:raspberry_pi}}
\label{tab:Raspberry_specs}
\begin{center}
\begin{tabularx}{\columnwidth}{|Y|Z|}
\hline
\textbf{CPU} & Arm Cortex-A76 de cuatro núcleos de 64 bits a 2.4GHz\\
\hline
\textbf{GPU} & VideoCore VII\\
\hline
\textbf{RAM} & 8GB LPDDR4X-4267 SDRAM\\
\hline
\textbf{Puertos} & 2xUSB3.0, 2xUSB2.0, Ethernet con soporte para PoE+, 2x4-lane MIPI\\
\hline
\textbf{Puertos de expansión} & header de 40 pines GPIO\\
\hline
\textbf{Memoria} & microSD con soporte para modo de lectura rápida SDR104\\
\hline
\end{tabularx}
\end{center}
\end{table}

Para la adquisición de datos, se ha instalado una tarjeta de lectura MCC 118, una placa tipo ``Hardware Attached on Top'' (HAT) desarrollada por Diligent específicamente para dispositivos Raspberry Pi. Esta placa está diseñada para medir voltajes simples en un rango de $\pm$10V con una precisión de 12 bits en hasta 8 canales simultáneos. También puede hacer barridos de lectura a velocidades de hasta 100kS/s, con un pin de trigger \cite{ref:MCC188_specs} que permite la sincronización de los barridos con la secuencia de conmutación del modulador de fase del instrumento. Aunque la placa no es capaz de trabajar por defecto con voltajes diferenciales, esta deficiencia fue solucionada por software a posteriori.

\begin{figure}[t]
\centering
\includegraphics[width=0.75\columnwidth]{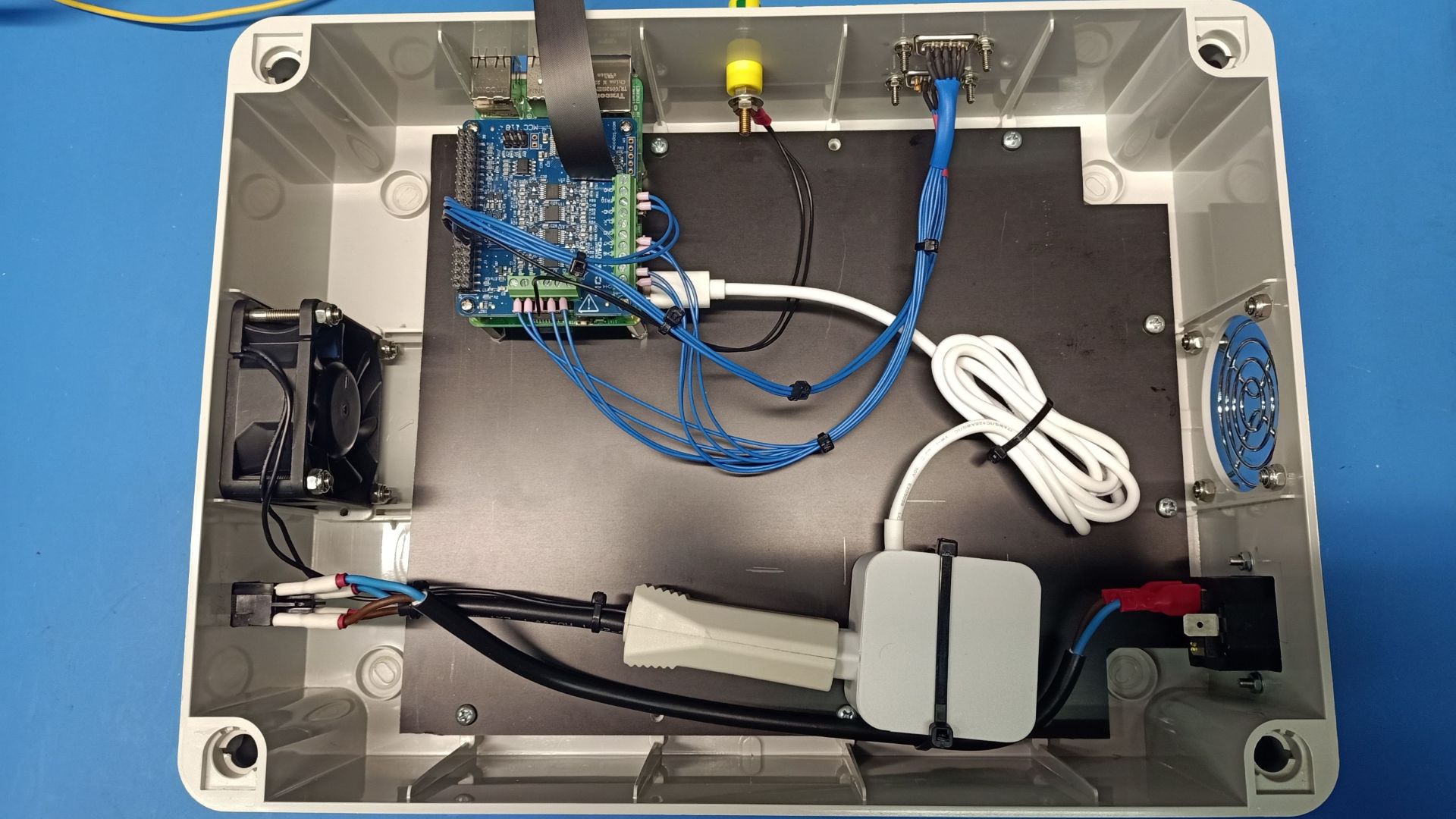}
\caption{Interior del calibrador TFGI QUIJOTE}
\label{fig:calibrator_interior}
\end{figure}

Para el control del modulador de fase, se han destinado pines libres del header de la Raspberry Pi. Tanto los canales de la MCC 118 como los pines de control han sido redirigidos a conectores sub-D de 9 pines en el panel trasero del dispositivo. También se ha añadido un puerto de referencia para poder igualar la tierra del instrumento a la del dispositivo y garantizar que las medidas sean correctas. Por último, se ha usado una pantalla táctil LCD DFR0678 de DFRobot con conector MIPI como interfaz humano-máquina para el dispositivo. En la fig. \ref{fig:calibrator_interior} puede observarse el interior del dispositivo finalizado.

\subsection{Software del dispositivo}

El software de funcionamiento del dispositivo sigue la misma filosofía de facilidad de uso y ampliación. El dispositivo funciona con sistema operativo Raspbian GNU/Linux, distribución de Linux propia de Raspberry Pi, open source y ampliamente usada y documentada en internet.

\begin{figure}[t]
\centering
\includegraphics[width=0.75\columnwidth]{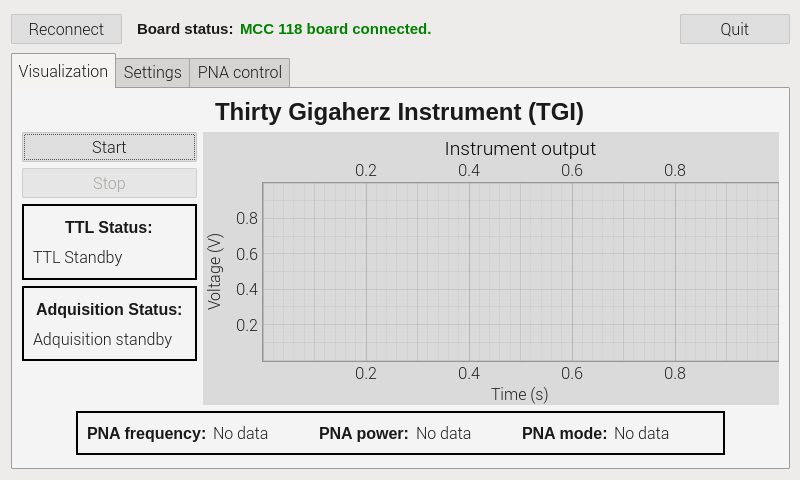}
\caption{Página principal de la interfaz del Calibrador TFGI QUIJOTE}
\label{fig:calibrator_interface}
\end{figure}

A su vez, el programa que sirve de interfaz de usuario del dispositivo está completamente implementado en Python, lenguaje ampliamente conocido y utilizado, lo que facilita su futura actualización y mejora. La comunicación con la MCC 118 se realiza usando la librería open source Daqhat, desarrollada y mantenida por el propio fabricante de la placa, mientras que la interfaz ha sido implementada usando PyQt5, ``wrapper'' para Python de la librería Qt. En la fig. \ref{fig:calibrator_interface} puede verse una captura de la página principal de la interfaz.

\section{Comparativa entre PXI y Calibrador TFGI QUIJOTE}

A continuación, se presenta una comparativa del Calibrador TFGI QUIJOTE con el PXI-1031DC.

\subsection{Capacidades}

En el apartado \ref{cap:calibrador}.\ref{subcap:calibrator_hardware} se describieron las características técnicas del hardware del Calibrador TFGI QUIJOTE. Si comparamos las capacidades de la Raspberry Pi con las del PXI-8186 que podemos ver en la tabla \ref{tab:PXI-8186_specs}, podemos comprobar que la Raspberry Pi posee características superiores a las del PXI, ya que no solo dispone de más potencia, sino que el uso de la GPIO de la Raspberry permite no necesitar de un módulo externo de E/S, como sí requiere el PXI.

\begin{table}[t]
\renewcommand{\arraystretch}{1.3}
\caption{Especificaciones PXI-8186 \cite{ref:PXI8186specs}}
\label{tab:PXI-8186_specs}
\begin{center}
\begin{tabularx}{\columnwidth}{|Y|Z|}
\hline
\textbf{CPU} & Pentium 4-M a 2.2GHz\\
\hline
\textbf{GPU} & Intel Extreme Graphics\\
\hline
\textbf{RAM} &  256 MB DDR\\
\hline
\textbf{Disco duro} & 30GB\\
\hline
\textbf{Puertos} & 2xpuertos seriales, 2xUSB2.0, Ethernet 100 BaseTX, interfaz GPIB (IEEE 488.2)\\
\hline
\end{tabularx}
\end{center}
\end{table}

Sin embargo, la tarjeta de lectura analógica PXI-4462 sí supone una ventaja importante del PXI con respecto al Calibrador TFGI QUIJOTE. Esta tarjeta también puede medir voltajes en el rango de $\pm$10V, y puede hacer barridos de lectura a velocidades de hasta 204.8kS/s con una precisión de 24 bits \cite{ref:PXI4462specs}, más del doble de lo que puede ofrecer la tarjeta MCC 118. Además, sus cuatro puertos están configurados para medir voltajes diferenciales, lo cual es relevante de cara al instrumento TFGI, ya que sus salidas son diferenciales.

\begin{figure}[t]
\centering
\includegraphics[width=0.85\columnwidth]{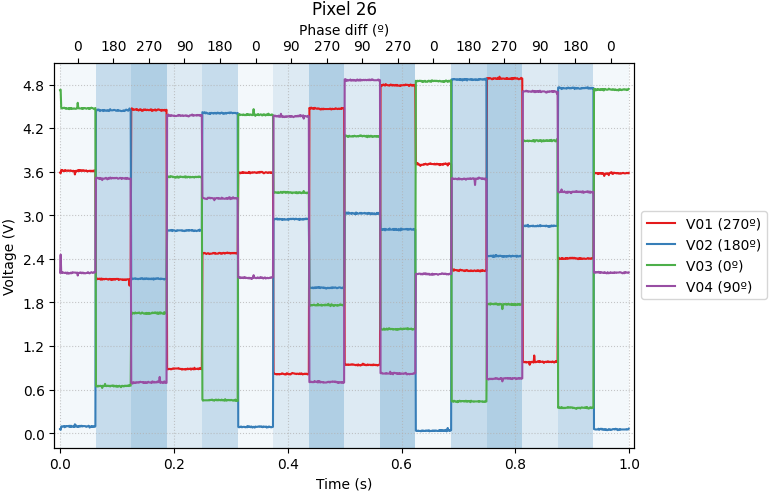}
\caption{Lectura de la salida del píxel 26 del TGI a una señal polarizada de 31 GHz y -50 dBm generada con el PNAx N5245B obtenida con datos capturados por el Calibrador TFGI QUIJOTE}
\label{fig:ejemplo_pixel}
\end{figure}

A pesar de lo anterior, la tarjeta MCC 118 mantiene la ventaja si el instrumento usa salidas de voltajes simples, donde la tarjeta contaría con hasta 8 puertos, frente a los 4 que mantendría el PXI-4462, ya que estos solo funcionan en modo diferencial. Además, dados los niveles de voltaje de salida con los que trabaja el instrumento normalmente, la diferencia en la precisión entre ambas tarjetas de lectura deja de resultar relevante. En la fig. \ref{fig:ejemplo_pixel} puede verse un ejemplo de señal caracterizada usando el Calibrador TFGI QUIJOTE para las cuatro salidas (Vd1, Vd2, Vd3 y Vd4) del instrumento TGI cuando recibe una señal con polarización Q pura a -50 dBm en la frecuencia central de la banda central del instrumento (31 GHz).

\begin{table*}[t]
\renewcommand{\arraystretch}{1.3}
\caption{Comparación de costes de un nuevo PXI y del Calibrador TFGI QUIJOTE\textsuperscript{1}}
\label{tab:Comparacion_PXI_Calibrador}
\begin{center}
\begin{tabular}{|c|c|c|c|}
\hline
\multicolumn{2}{|c|}{\textbf{PXI}} & \multicolumn{2}{c|}{\textbf{Calibrador}} \\
\hline
\textbf{Componente} & \textbf{Precio} & \textbf{Componente} & \textbf{Precio} \\
\hline
Núcleo PXI-8822 & 3698 € & Núcleo Raspberry Pi 5 & 106.48€ \\
\hline
Módulo lectura analógica PXIe-4464 &  7993 € & Módulos MCC 118 & 84.77€ \\
\hline
Módulo E/S PXI-6509 &  937 € & Cable de alimentación Raspberry & 14.97 € \\
\hline
Chasis NI PXIe-1071 & 1957 € & Carcasa y resto de componentes & 95€ \\
\hline
\textbf{Subtotal} & \textbf{14585 €} & \textbf{Subtotal} & \textbf{304.18 €}\\
\hline
\end{tabular}
\end{center}
\begin{center}
\parbox{0.65\textwidth}{\footnotesize{\textsuperscript{1}Precios de referencia promedios en proveedores oficiales a marzo de 2026}}
\end{center}
\end{table*}

\subsection{Facilidad de uso}

Como ya se mencionó en el apartado \ref{cap:introduction}, el controlador PXI-8186 del PXI-1031DC funciona con Windows XP. Esta es una versión con drivers y software privativos de National Instruments y no puede ser actualizado tanto por este motivo como por la antigüedad del dispositivo, lo que limita las opciones de desarrollo de aplicaciones para la caracterización del instrumento al uso del programa LabVIEW 2012, el cual no solo es una versión antigua del programa, sino que también se trata de un programa con una pronunciada curva de aprendizaje y no demasiado conocido fuera de ciertos campos.

En cambio, el Calibrador TFGI QUIJOTE, al estar implementado sobre un dispositivo basado en Linux, todo su software es libre y fácilmente actualizable, además de que es compatible con tecnologías y lenguajes modernos como Python, lo que facilita las modificaciones y futuras mejoras del dispositivo. Además, la interfaz implementada actualmente en el dispositivo se diseñó pensando en la sencillez y facilidad de uso, permitiendo que incluso usuarios no formados en el diseño del equipo puedan usarlo sin complicaciones.

El peso es otro punto a favor para el Calibrador TFGI QUIJOTE de cara a facilitar su manejo. Con todos sus módulos instalados, el peso total del PXI-1031DC supera los 6 kg \cite{ref:PXI8186specs}\cite{ref:PXI4462specs}\cite{ref:PXI1031specs}, lo cual dificulta su manejo cómodo a la hora de tener que desplazarlo al lugar del instrumento. Por otro lado, una vez finalizado el ensamblaje, se ha estimado el peso del Calibrador es de menos de 1 kg, lo que permite su manejo cómodo con una sola mano. Además, su pantalla táctil permite que, en caso de necesidad, no sea necesario ningún periférico extra para la visualización y control, mientras que el PXI requiere el uso de una pantalla, teclado y ratón para su uso local. 

\subsection{Coste}

Una de las mayores ventajas del Calibrador TFGI QUIJOTE con respecto al PXI-1031DC radica en el coste de estos dispositivos. Muchas de las taras y deficiencias mencionadas del PXI con respecto al Calibrador pueden ser solucionadas con la adquisición de una versión más moderna de este. Sin embargo, el alto coste de estos dispositivos lo hace poco atractivo frente al Calibrador. Como puede observarse en la tabla \ref{tab:Comparacion_PXI_Calibrador}, el coste de un solo componente del PXI ya es superior al coste total del Calibrador TFGI QUIJOTE, lo que hace que el dispositivo resulte mucho más competitivo económicamente que el PXI.

\section{Conclusiones}

El Calibrador TFGI QUIJOTE se muestra como una alternativa sólida a los sistemas PXI tradicionales para su uso en la caracterización y calibración de instrumentos de radioastronomía como el TFGI. Su diseño sencillo y modular, basado en hardware y software open source y de uso ampliamente extendido tanto en ``maker'' como profesional, garantiza no solo su facilidad de utilización y la calidad de los resultados obtenidos, sino que permite su fácil adaptación a otros sistemas y la implementación de mejoras.

Actualmente, tras esta primera versión inicial, ya se encuentran planificadas diversas mejoras, como un controlador para rotores que permita el control de la rotación de las bocinas usadas como emisores durante los procedimientos de caracterizado y calibrado del instrumento, lo que permitiría controlar la polarización de la señal de entrada al instrumento, además de la implementación de una opción para la realización automática de barridos tanto en frecuencia como en potencia.


\section*{Agradecimientos}

Los autores agradecen la financiación de este trabajo, obtenida con los proyectos de investigación con referencia PID2022-139223OB-C21 y con fondos europeos ``European Union NextGenerationEU/PRTR''. Se agradece también la financiación parcial del Ministerio de Ciencia, Innovación y Universidades, proyectos: AYA2017-84185-P, IACA15-BE-3707, EQC2018-004918-P y el Acuerdo FEDER INSIDE-OOCC (ICTS-2019-03-IAC-12). También se agradece el apoyo de los programas Servero Ochoa SEV-2015-0548 y CEX2019-000920-S. Agradecimientos también al apoyo y soporte obtenido por parte del Departamento de Ingeniería de Comunicaciones de la Universidad de Cantabria (DICOM-UC). Finalmente, se agradece a Airam Marcos-Caballero por su ayuda en la redacción con \LaTeX.


\bibliography{bibliography_spanish_version}